\documentclass[11pt,showpacs,preprintnumbers,amsmath,amssymb,prd,nofootinbib,superscriptaddress]{revtex4-2}

\usepackage{dcolumn}% Align table columns on decimal point
\usepackage{bm}% bold math
\usepackage{ifpdf}
\usepackage{hyperref}
\usepackage{xcolor,color,graphicx,graphics,physics}
\usepackage[spanish,english]{babel}%, portuguese
\usepackage[latin1]{inputenc}
\usepackage[OT1]{fontenc}
\usepackage{latexsym,amssymb,amsmath,amsfonts, slashed,cancel}
\usepackage{makeidx}
\usepackage{epsfig,subfigure}
\usepackage{natbib}
\usepackage{epstopdf}
\usepackage{mathrsfs}
\usepackage{hyperref}%\usepackage[colorlinks=true,linkcolor=blue]{hyperref}
\hypersetup{colorlinks=true, linkcolor=blue, citecolor=blue, urlcolor=blue}
\usepackage{enumerate}

\usepackage{fixmath}

\everymath{\displaystyle}
\usepackage{graphicx}

\usepackage[T1]{fontenc}
\usepackage{amsmath}
\usepackage{amssymb}
\usepackage{graphicx}
\usepackage{xcolor}

\newcommand{\bea}{\begin{eqnarray}}
\newcommand{\eea}{\end{eqnarray}}

\newcommand{\orcid}[1]{\href{https://orcid.org/#1}{\includegraphics[width=10pt]{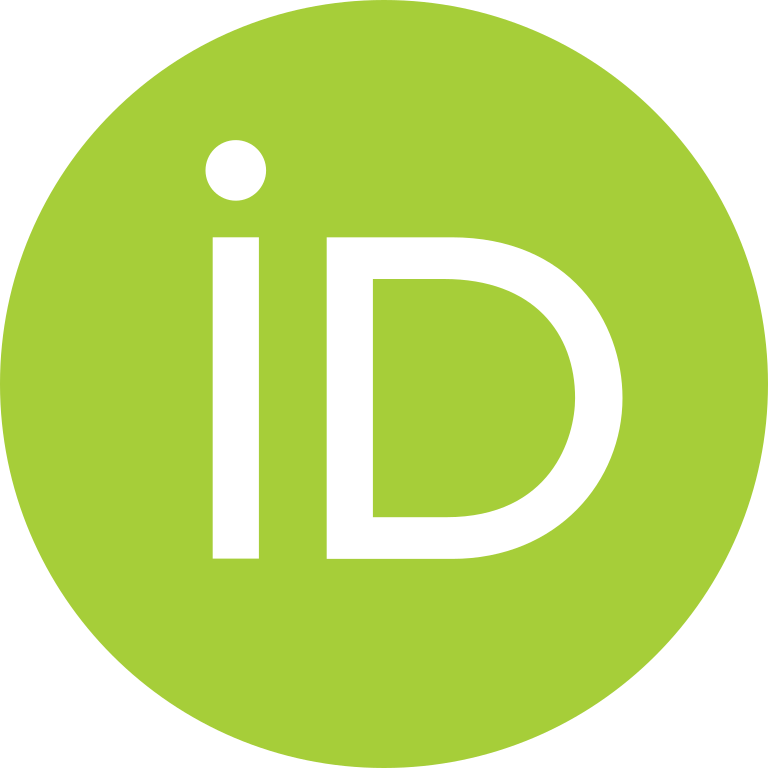}}}

\begin{document}

%\title{Chiral lepton pair production with EDM from a Electron-Positron annihilation under a strong magnetic field and Lorentz symmetry breakdown}

%\title{Leptons pair production from a electron-positron annihilation: EDM, magnetic field and Lorentz violation correction}

\title{Exploring the impact of magnetic fields, Lorentz violation and EDM on $e^+ e^-\rightarrow l^+ l^-$ scattering}

\author{D. S. Cabral  \orcid{0000-0002-7086-5582}}
\email{danielcabral@fisica.ufmt.br}
\affiliation{Programa de P\'{o}s-Gradua\c{c}\~{a}o em F\'{\i}sica, Instituto de F\'{\i}sica,\\ 
Universidade Federal de Mato Grosso, Cuiab\'{a}, Brasil}

\author{A. F. Santos \orcid{0000-0002-2505-5273}}
\email{alesandroferreira@fisica.ufmt.br}
\affiliation{Programa de P\'{o}s-Gradua\c{c}\~{a}o em F\'{\i}sica, Instituto de F\'{\i}sica,\\ 
Universidade Federal de Mato Grosso, Cuiab\'{a}, Brasil}

\begin{abstract}

This paper explores the annihilation process of an electron-positron pair into a heavier lepton-antilepton pair, taking into account the presence of an external classical magnetic field. Additionally, it investigates corrections arising from the breakdown of Lorentz symmetries and the existence of Electric Dipole Moments (EDM) for leptons.  Graphics are constructed to illustrate the influence of EDM and Lorentz violation on the cross section. Strong magnetic field limit is analyzed.  Furthermore, the investigation utilizes experimental data on EDM values to explore the upper limits of the Lorentz violation parameters. 

\end{abstract}

\maketitle

\section{Introduction}

Many works deal with the study of Quantum Electrodynamics (QED) processes involving Lorentz violation (LV). This occurs when there is a breakdown in Lorentz symmetry between particle and frame transformations. In such cases, researchers often introduce a constant background tensor into the theory, thereby selecting preferential directions \cite{colladayredef, kostelecky1, kostelecky2, ale1, daniel1}. The upper limits for the components of these tensors have been obtained in various situations through numerous experimental works. Many of these experimental constraints are compiled in tables, which are regularly updated \cite{tableviolation}. The electron sector comprises the majority of limits for Lorentz and CPT violation parameters among leptons. This can be attributed, in part, to the electron's stability compared to other leptons. Additionally, the focus of many studies on electron behavior contributes to the abundance of experimental data and constraints in this sector \cite{electron1, electron2, electron3}.  

Another ingredient that can be taken into account in the usual QED theory calculations is the existence of Electric Dipole Moment (EDM) for leptons. This feature can arise in the theory through a sensitive analysis of Electromagnetic Form Factors (EFF). These quantities for hadrons arise from quantum corrections of the interaction current as Fourier transforms of magnetization and charge distributions, which can be calculated from loops in the QED theory \cite{protonformfactor, formfactorbook, protonformfactor2}. These concepts can be extended to any spin-1/2 particle in such a way that these quantities become functions of the momentum transfer \cite{formfactors}. Some of these characteristics can be found when we study them together with symmetry breakdowns and violations \cite{edmexpression}.

Looking exclusively at the EDM contribution, it is known that the existence of this quantity violates T and, consequently, CP symmetries \cite{protonformfactor2, edm01}. For this reason, in the search for physics beyond the standard model, there has been extensive research into the non-zero values of the EDM for leptonic particles and efforts to obtain upper limits for this quantity, primarily for the Muon and Tau leptons. This has led to numerous experimental tests and, consequently, multiple results regarding the true value of this form factor \cite{edm00, edm01, edm02, edm03, edm04, edm05, edm06}. In terms of symmetry breakdowns, some studies are dedicated to analyzing the form factor from the perspective of Lorentz violation. They conclude that this quantity may experience enhancement at very high energies, potentially reaching magnitudes as large as $10^{-14} e\cdot \text{cm}$\footnote{Here $e$ is the electronic charge and $e\cdot \text{cm}=1.608\times 10^{-19}\, C.cm \,(Coulomb.centimeter)$.} \cite{process}. The upper limit of the Electron EDM, in one of the most recent updates, is set at $|\varrho_e|<1.1\times10^{-29}e\cdot\text{cm}$ with 90\% confidence \cite{electronedm}. Despite this value being very small, one might expect that this feature can be measured for heavier leptons, since the electric dipole moments for these particles can be up to a million times greater than that of the Electron \cite{edm04}. This can be observed through limits imposed on the Muon \cite{edm03} and the Tau \cite{edm05}.

For certain energy ranges, the lepton EDM contribution is comparable to some weak effects \cite{edmweak}. Therefore, measuring non-zero form factors for point-like particles as desired directly impacts the limits and range of validity of the standard model theory. An analysis of the form factors using the formalism of extended QED allows for the derivation of an analytic expression of the EFF, which can generate the lepton EDM from the LV tensor $d_{\mu\nu}$ \cite{edmexpression}. Following this approach, one can use measured limits to impose constraints on the Lorentz parameters for the Muon and Tau leptons. Thus, with enhancement due to a high-energy scenario, either the EDM or LV effects could be observed. The motivation behind adding these features lies in the desire to model certain aspects of the early universe, which is characterized by very high energies and strong external fields. By incorporating this last feature, the cross-section can attain values that are greater than without the presence of an external field \cite{process}. To observe these effects, many studies are dedicated to obtaining modifications of scatterings due to the presence of a constant external magnetic field, particularly in a specific direction, such as the $z$-coordinate, for example \cite{danielmag, mag01, mag02, mag03}.

This paper is organized as follows: Sec. \ref{secformfactor} explains the meaning of electromagnetic form factors. Sec. \ref{secdiracfield} addresses the formulation of the LV-Dirac field subject to an external magnetic field. Subsections \ref{secparticle}  focus on obtaining the full form of the spinors for particles and antiparticles in the chiral representation, respectively. Subsection \ref{secstrongerfields} discusses the approximation of external fields with very high magnitudes. Finally, Sec. \ref{secscattering} studies $e^{-}e^{+}\to l^{-}l^{+}$ scattering, taking into consideration the EDM and LV vertices in addition to the usual QED ones. An estimation of the Lorentz violating parameter is made based on the upper limits of measured EDMs for muonic and tauonic particles. These results are represented graphically and compared with results regarding the electron. Conclusions are presented in Sec. \ref{secconclusion}.

\section{Electromagnetic form factors and EDM}\label{secformfactor}

In this section, we derive the electromagnetic form factors by utilizing the Dirac current. It is known that the Dirac 4-current $J^\mu$ must satisfy the continuity equation
\begin{eqnarray}
\partial_\mu J^\mu=\partial_0\rho +\vec{\nabla}\cdot\vec{J}=0.
\end{eqnarray}
By introducing the change $\vec{J}\to\vec{J}+\vec{\nabla}\times\vec{\mathcal{J}}$, this new current can be expressed in a form that preserves the continuity equation. In the covariant form, the current becomes $J^\mu\to\Theta^\mu=\left(  \rho,\vec{J}+\vec{\nabla}\times\vec{\mathcal{J}}\right)$. In terms of the Dirac field $\psi$ we have \cite{paulicurrent}
\begin{eqnarray}
\vec{J}\to\vec{J}+\vec{\nabla}\times\vec{\mathcal{J}}=-\frac{i}{2m}\left[\bar{\psi}\left(\vec{\nabla}\psi\right)-\left(\vec{\nabla}\bar{\psi}\right)\psi\right]+\vec{\nabla}\times\vec{\mathcal{J}}
\end{eqnarray}
which implies that the modified current $\Theta^\mu(x)$ can be a combination of a certain set of gamma matrices. 

To proceed, our task is to generate a new current immediately after encountering $\vec{\mathcal{J}}$, achieved by combining Dirac $4\times4$ matrices. Our aim is to investigate the behavior of the quantity $\bra{p}\Theta^\mu(x)\ket{p^\prime}=e^{-i(p^\prime-p)x}\bra{p}\Theta^\mu(0)\ket{p^\prime}$. Furthermore, these combinations of gamma matrices must adhere to specific conditions \cite{formfactors}, i.e., (i)  Lorentz covariance: $\Theta_\mu(x)$ transforms as a 4-vector under Lorentz transformations. (ii) Hermiticity: $\Theta_\mu^\dagger(x)=\Theta_\mu(x)$ behaves like a Hermitian operator. (iii) Gauge invariance: $\partial_\mu\Theta^\mu(x)=0$, charge needs to be conserved. These requirements can be translated into momentum space, that is,
\begin{eqnarray}
\bra{p}\Theta^\mu(x)\ket{p^\prime}=e^{-i(p^\prime-p)x}\bar{u}(p)O^\mu(q)u(p^\prime),
\end{eqnarray}
where $q^\mu=p^{\prime\mu}-p^\mu$ is the momentum transfer. The characteristics of Lorentz covariance, hermiticity and gauge invariance can be rewritten to the momentum-space operator $O^\mu(q)$. Then, we can propose some candidates for that set of matrices. The first options are $\{\mathbb{I},\gamma^5,\sigma_{\mu\nu}\}$, along with combinations involving $\{p^\mu,p^{\prime\mu}\}$, and those formed by the matrices $\{\gamma^\mu,\gamma^5\}$ in conjunction with the Levi-Civita tensor $\{\epsilon^{\mu\nu\alpha\beta}\}$.

Using some gamma matrix identities and the Gordon decomposition, the following contributions can be written
\begin{eqnarray}
\bar{u}(p)O_{\mu}(q)u(p^{\prime})=\bar{u}(p)\left\{F_1(q^2)\gamma_\mu+i\frac{F_2(q^2)}{2m}\sigma_{\mu\nu}q^\nu+\frac{F_3(q^2)}{2m}\sigma_{\mu\nu}q^\nu\gamma^5
\right\}u(p^{\prime}).\label{eq78}
\end{eqnarray}
These form factors are defined, at zero recoil $q^2=0$, as
\begin{eqnarray}
F_1(0)=\text{Charge};\quad\quad \frac{1}{2m}\left[F_1(0)+F_2(0)\right]=\text{Magnetic moment};\quad\quad -\frac{1}{2m}F_3(0)=\text{EDM}.\label{eq79}
\end{eqnarray}
This definition can be derived from an analysis of the non-relativistic limit of an interaction Hamiltonian describing a particle in an electromagnetic field, as demonstrated by \cite{formfactors}.

The charge form factor is a well-established term in QED theory. It becomes evident when coupling Eq. (\ref{eq78}) with a photon field, leading to $\bar{u}(p)O_{\mu}(q)A^\mu u(p^{\prime})=\bar{u}(p)F_1(q^2)\gamma_\mu A^\mu u(p^\prime)$, which implies an electric charge of $F_1(q^2)=-e$. The second contribution term of Eq.  (\ref{eq79}) pertains to the magnetic moment of the electron and other heavier leptons, which has been extensively researched due to its deviation from traditional Dirac theory \cite{anomalous1, anomalous2, anomalous3, anomalous4, anomalous5, anomalous6, anomalous7}. One of the most accurate and up-to-date values for the anomalous magnetic moment of the muon is provided by \cite{anomalous4}, yielding $a=116592061(41)\times10^{-11}$. The third contribution is the form factor that can generate the leptonic EDM. An expression for the enhancement of the heavier charged leptons at high energies, based on Lorentz-violating physics, was derived by \cite{edmexpression} and yields
\begin{eqnarray}
F_3(0)=\frac{275\alpha}{18\pi}\frac{p^\mu d_{\mu\nu}^{(S)}p^\nu}{m^2}, \label{eq82}
\end{eqnarray}
where $p^\mu$ and $m$ denote the momentum and mass of the particle, respectively, and the superscript $(S)$ represents the symmetric part of the Lorentz violation coefficient $d_{\mu\nu}$.

In addition, certain texts introduce an additional form factor known as the Anapole moment term, as proposed by Zel'dovich \cite{zel}. While conserving CP-symmetry, it violates parity. This quantity has been calculated for the electron within the context of electroweak processes \cite{anapoleweak}. It relates to how matter couples with electric and magnetic fields and arises in scattering processes involving fermions and virtual photons, particularly when the exchanged photon is off-shell \cite{formfactors}. From this perspective, the electromagnetic form factors work as quantum corrections for composite particles. In Eq. (\ref{eq78}), we observe that the first term (related with the electric charge) implies that the fermion is a ``Bare particle'', while the subsequent terms emerge as ``self-interactions''.

The following section addresses the Dirac field in the presence of constant background tensors that break Lorentz symmetry.

\section{The Dirac field}\label{secdiracfield}

This section presents the Dirac Lagrangian incorporating Lorentz-violating terms. Additionally, the Dirac equation is solved considering the presence of an external electromagnetic field. The extended Lagrangian for QED is provided as
\begin{eqnarray}
\mathcal{L}=\bar{\psi}\left[i\left(\gamma^\mu+c^{\nu\mu}\gamma_\nu+d^{\nu\mu}\gamma^5\gamma_\nu+e^\mu+if^\mu\gamma^5+\frac{1}{2}g^{\lambda\nu\mu}\sigma_{\lambda\nu}\right)\mathcal{D}_\mu-m\right]\psi,
\end{eqnarray}
where $c^{\nu\mu}$, $d^{\nu\mu}$, $e^{\mu}$, $f^{\mu}$, and $g^{\lambda\nu\mu}$ are real quantities that introduce Lorentz violation into the problem \cite{violacao1kostelecky}. Due to its axial nature, when compared with the modified current $\Theta^\mu$, the only coefficients that can give rise to the EDM in our problem are $f^\mu$ and $d^{\nu\mu}$. However, since the former can be moved to the $c_{\mu\nu}$ sector, by a field redefinition \cite{eliminatef}, we will consider only the coefficient $d^{\nu\mu}$ for our purposes. Then the Lagrangian becomes
\begin{eqnarray}
\mathcal{L}&=& \bar{\psi}\left[i\left(\gamma^\mu+d^{\nu\mu}\gamma^5\gamma_\nu\right)\mathcal{D}_\mu-m\right]\psi\nonumber\\
&=&\bar{\psi}\,i\left[\left(\partial_\mu+d_{\mu\nu}\gamma^5\partial^\nu\right)-ie\left(A_\mu+d_{\mu\nu}\gamma^5 A^\nu\right)\right]\gamma^\mu\psi-m\bar{\psi}\psi.\label{eq76}
\end{eqnarray}
Here, $\mathcal{D}_{\mu}=\partial_\mu+iqA_\mu$ with $q=-e$ being the electron charge and $A_\mu$ representing the gauge field. This equation implies a "new" free Dirac equation that can couple to the photonic field in two different vertices, namely $ie\gamma_\mu$ and $d_{\mu\nu}\gamma^5\gamma^\nu$.

If we want to study a problem with an electromagnetic external field in addition to those describing the interacting photons, we need to express $A^\mu=A_D^\mu+A_B^\mu$, where $A_D^\mu$ and $A_B^\mu$ represent the dynamical and external parts of the potential, respectively. If this field is a classical external magnetic field constant in the $z$-direction, it can be written as $A_B^\mu=(0,-yB,0,0)$, such that $\vec{B}=B\hat{z}$.

In this formalism, the new field is determined by solving the modified Dirac equation
 \begin{eqnarray}
 i(g_{\mu\nu}+d_{\mu\nu}\gamma^5)\gamma^{\mu}\partial^\nu\psi+e(g_{\mu\nu}+d_{\mu\nu}\gamma^5)\gamma^\mu A^\nu_B\psi-m\psi=0,\label{eq01}
 \end{eqnarray}
where the chiral representation is used. In this representation we have
\begin{eqnarray}
\gamma^0=\begin{pmatrix}
0 & 1\\
1 & 0
\end{pmatrix};\quad\quad \gamma^k=\begin{pmatrix}
0 & \sigma^k\\
-\sigma^k & 0
\end{pmatrix};\quad\quad \gamma^5=\begin{pmatrix}
-1 & 0 \\
0 & 1
\end{pmatrix}.
\end{eqnarray}
The terms $0$ and $1$ represent $2\times2$ matrix blocks that are the null and identity matrices, respectively, while $\sigma^k$ denotes the $k$-th Pauli matrix. It's worth noting from Eq. (\ref{eq76}) that it's necessary to separate the electromagnetic vector into its external and photonic components. However, the photonic component is relevant only at the vertex of the reaction, while the \text{external} component modifies the standard Dirac equation, resulting in the new equation shown in Eq. (\ref{eq01}), as demonstrated in \cite{danielmag}.

\subsection{Particle and Anti-particle solutions}\label{secparticle}

The objective here is to solve the modified Dirac equation and find solutions for both particles and antiparticles. For positive energy spinors, we can consider the solution as
\begin{eqnarray}
\psi=e^{-iEt}\begin{pmatrix}
\psi_L\\\psi_R
\end{pmatrix}.
\end{eqnarray}
To avoid non-standard time derivatives, we have two options: either perform a field redefinition as suggested in \cite{colladayredef}, or set $d^{\mu0}=d^{0\mu}=0$, thereby transforming Eq. (\ref{eq01}) into
\begin{eqnarray}
E\gamma^0\psi+g_{jk}\gamma^j(i\partial^k+eA^k)\psi+d_{jk}\gamma^5\gamma^j(i\partial^k+eA^k)\psi-m\psi=0.\label{eq84}
\end{eqnarray}
Alternatively, after expressing it in matrix block form, we obtain the following equations for the bi-spinors
\begin{eqnarray}
\begin{cases}
E\psi_R+\left(g_{jk}-d_{jk}\right)\sigma^{j}\left(i\partial^k+eA^k\right)\psi_R-m\psi_L=0 & \text{(a)}\\
E\psi_L-\left(g_{jk}+d_{jk}\right)\sigma^{j}\left(i\partial^k+eA^k\right)\psi_L-m\psi_R=0. & \text{(b)}
\end{cases}\label{eq02}
\end{eqnarray}

From Eq.(\ref{eq02}(a)) we get
\begin{eqnarray}
\psi_R=\left[E+\left(g_{jk}-d_{jk}\right)\sigma^{j}\left(i\partial^k+eA^k\right)\right]^{-1}m\psi_L.\label{14}
\end{eqnarray}
Using Eq. (\ref{14}) in Eq.(\ref{eq02}(b)), the equation of motion for 2-spinor $\psi_L$ is expressed as
\begin{eqnarray}
\left[E+\left(g_{jk}-d_{jk}\right)\sigma^{j}\left(i\partial^k+eA^k\right)\right]\left[E-\left(g_{jk}+d_{jk}\right)\sigma^{j}\left(i\partial^k+eA^k\right)\right]\psi_L-m^2\psi_L=0.\label{eq46}
\end{eqnarray}

At this moment, it is crucial to conduct a preliminary analysis of the LV-coefficient and delineate its key characteristics. According to classical principles, the momentum $\vec{\pi}$ of a charged particle in the presence of an external field is typically expressed as $\vec{\pi}=\vec{p}-e\Vec{A}$, with the total energy depending on $\Vec{\sigma}\cdot\Vec{\pi}$,  representing the projection of the total momentum onto the spin direction. However, upon examining Eq. (\ref{eq84}) and, consequently, Eq. (\ref{eq46}), we observe that this same projection emerges in a modified form, denoted as $-(g_{jk}-d_{jk})\sigma^j\left(-i\partial^k-eA^k\right)$. This observation reveals that the LV-term introduces an off-diagonal coupling between the total momentum and the spin components, alongside a modification in the coupling strength of the diagonal contributions. So, if we choose $d_{\mu\nu}$ to be non-vanishing only for $d_{33}$, we obtain a correction to the theory solely for the projection of $\pi_z$ into the $\sigma_z$ direction, i.e., the direction in which the field lies. In other words, we aim to analyze the effects of incorporating this specific component
\begin{eqnarray}
    d_{\mu\nu}=\begin{pmatrix}
        0 & 0 & 0 & 0\\
        0 & 0 & 0 & 0\\
        0 & 0 & 0 & 0\\
        0 & 0 & 0 & d_{33}
    \end{pmatrix}.
\end{eqnarray}

\subsubsection{Left-handed solutions}

For what follows, we have to consider the full form of the left-handed bi-spinor, which we take as $\psi_L=\begin{pmatrix}
\phi_1 \\ \phi_2
\end{pmatrix}$. Additionally, we suppose the following solution
\begin{eqnarray}
\phi_1=e^{i\vec{p}\cdot\vec{r}_{\cancel{y}}}
f_1(y);\quad\quad \phi_2=e^{i\vec{p}\cdot\vec{r}_{\cancel{y}}}f_2(y),
\end{eqnarray}
where $r_{\cancel{y}}$ refers to the position with $y$-component set to zero, that is, $\vec{p}\cdot\vec{r}_{\cancel{y}}=p_xx+p_zz$ as the formalism developed by \cite{magfield,magfieldthesis}. In this way, the application of the sigma spin operators over $\psi_L$ (the exponential is suppressed for shorter notation) leads to
\begin{gather}
\sigma^1\psi_L=\begin{pmatrix}
f_2(y)\\f_1(y)
\end{pmatrix};\quad\quad \sigma^2\psi_L=\begin{pmatrix}
-if_2(y)\\if_1(y)
\end{pmatrix};\quad\quad \sigma^3\psi_L=\begin{pmatrix}
f_1(y)\\-f_2(y)
\end{pmatrix}.
\end{gather}
This way, the problem, when returning to Eq (\ref{eq46}), will be
\begin{eqnarray}
\begin{cases}
\left(a_0+a_1y+a_2y^2\right)f_1+\frac{d^2f_1}{dy^2}+\left(a_3+a_4y\right)f_2+a_5\frac{df_2}{dy}=0,\\
\left(b_0+b_1y+b_2y^2\right)f_2+\frac{d^2f_2}{dy^2}+\left(b_3+b_4y\right)f_1+b_5\frac{df_1}{dy}=0,\label{eq49}
\end{cases}
\end{eqnarray}
where the constants are defined as
\begin{eqnarray}
a_0&=&E^2-m^2-2Ep_zd_{33}-\left(p_x^2+p_z^2+eB\right);\nonumber\\
a_1&=&2eBp_x;\quad\quad a_2=-e^2B^2;\quad\quad a_3=-2p_xp_zd_{33};\nonumber\\
a_4&=&2eBp_zd_{33};\quad\quad a_5=2p_zd_{33}
\end{eqnarray}
and
\begin{eqnarray}
b_0&=&E^2-m^2+2Ep_zd_{33}-\left(p_x^2+p_z^2-eB\right);\nonumber\\
b_1&=&2eBp_x;\quad\quad b_2=-e^2B^2;\quad\quad b_3=2p_xp_zd_{33};\nonumber\\
b_4&=&-2eBp_zd_{33};\quad\quad b_5=2p_zd_{33}.
\end{eqnarray}
It is important to note that we have to solve a system with two differential equations for $f_1(y)$ and $f_2(y)$ which, due to the nature of the LV term contribution, are coupled. Therefore, to obtain the solutions for $\psi_L$ and, consequently, for $\psi_R$, it is necessary to solve Eq. (\ref{eq49}) for both functions simultaneously.

For simplicity, let us rewrite Eq. (\ref{eq49}) in terms of a redefinition of the linear differential operators, i.e.,
\begin{eqnarray}
\begin{cases}
\mathbb{L}_1\left\{f_1\right\}-d_{33}\mathbb{L}_a\{f_2\}=0,\\
\mathbb{L}_2\{f_2\}-d_{33}\mathbb{L}_b\{f_1\}=0,
\end{cases}\label{eq50}
\end{eqnarray}
where, $\mathbb{L}_{i}$, for $i=1,2,a,b$, are differential operators defined as
\begin{eqnarray}
\mathbb{L}_1\equiv&\left(a_0+a_1y+a_2y^2\right)+\frac{d^2}{dy^2};\quad\quad\mathbb{L}_a\equiv&\left(2p_xp_z-2eBp_zy\right)-2p_z\frac{d}{dy};\nonumber\\
\mathbb{L}_2\equiv&\left(b_0+b_1y+b_2y^2\right)+\frac{d^2}{dy^2};\quad\quad\mathbb{L}_b\equiv&\left(-2p_xp_z+2eBp_zy\right)-2p_z\frac{d}{dy}.\label{eq51}
\end{eqnarray}

To solve these equations, let's assume a solution of the type
\begin{eqnarray}
f_1=\sum_{k=0}^{\infty} \left(d_{33}\right)^kF_k\quad\quad\text{and}\quad\quad f_2=\sum_{k=0}^{\infty}\left(d_{33}\right)^kG_k.\label{eq52}
\end{eqnarray}
This approach is called the perturbation method and yields the final functions $f_1$ and $f_2$ as perturbative corrections of the functions $F_0(y)$ and $G_0(y)$ in terms of a power series of the $d$-coefficients \cite{perturbation1, perturbation2, perturbation3}.

Taking equations (\ref{eq52}) into their explicit forms, we have
\begin{eqnarray}
f_1(y)=F_0(y)+d_{33}F_1(y)+d_{33}^2F_2(y)+d_{33}^3F_3(y)+\mathcal{O}(d_{33}^4)\label{eq59}
\end{eqnarray}
and
\begin{eqnarray}
f_2(y)=G_0(y)+d_{33}G_1(y)+d_{33}^2G_2(y)+d_{33}^3G_3(y)+\mathcal{O}(d_{33}^4).\label{eq60}
\end{eqnarray}
Since the LV term is small, we will only consider terms up to the lowest order. In this way, substituting Eqs. (\ref{eq52}) into Eqs. (\ref{eq50}), we get\begin{eqnarray}
0&=&\mathbb{L}_1\left\{F_0\right\}+d_{33}\mathbb{L}_1\left\{F_1\right\}- d_{33}\mathbb{L}_a\left\{G_0\right\}
\end{eqnarray}
and
\begin{eqnarray}
0&=&\mathbb{L}_2\left\{G_0\right\}+d_{33}\mathbb{L}_2\left\{G_1\right\}-d_{33}\mathbb{L}_b\left\{F_0\right\}.
\end{eqnarray}

So, as we are supposing $d_{33}$ is a small but non-zero parameter, we have two conditions up to first order,
\begin{eqnarray}
\begin{cases}
\mathbb{L}_1\{F_0(y)\}=0\quad\quad\text{(A1)};\\
\mathbb{L}_2\{G_0(y)\}=0\quad\quad\text{(A2)};
\end{cases}
\quad\quad 
\begin{cases}
\mathbb{L}_1\left\{F_1\right\}-\mathbb{L}_a\{G_0\}=0;\quad\quad\text{(A3)}\\
\mathbb{L}_2\left\{G_1\right\}-\mathbb{L}_b\{F_0\}=0.\quad\quad\text{(A4)}
\end{cases}\label{eq54}
\end{eqnarray}

For Eq. (\ref{eq54}A1) with the definitions given by Eq. (\ref{eq51}), the solution becomes
\begin{eqnarray}
F_0(y)=e^{-\frac{1}{2eB}\left(eBy-p_x\right)^2}H_n\left(\frac{eBy-p_x}{\sqrt{eB}}\right),\label{eq28}
\end{eqnarray}
where $H_n(x)$ is the $n-$th Hermite polynomial, with the condition to be satisfied given by
\begin{eqnarray}
E^2-m^2-2Ep_zd_{33}-p_z^2&=&2(n+1)eB,
\end{eqnarray}
which are the energy eigenvalues. This solution was expected since it is analogous to Landau's problem.

For Eq. (\ref{eq54}A2), an analogous result is obtained
\begin{eqnarray}
G_0(y)=e^{-\frac{1}{2eB}\left(eBy-p_x\right)^2}H_n\left(\frac{eBy-p_x}{\sqrt{eB}}\right)
\end{eqnarray}
whose energy levels are given as
\begin{eqnarray}
E^2-m^2+2Ep_zd_{33}-p_z^2&=&2neB.
\end{eqnarray}

To simplify what follows, let's define a special function $I_n(\upsilon)$ as
\begin{eqnarray}
I_n(\upsilon)=N_ne^{-\frac{{\upsilon}^2}{2}}H_n\left(\upsilon\right),
\end{eqnarray}
where $\upsilon=\frac{eBy-p_x}{\sqrt{eB}}$ is a new variable and
\begin{eqnarray}
N_n=\left(\frac{\sqrt{eB}}{2^nn!\sqrt{\pi}}\right)^{\frac{1}{2}}
\end{eqnarray}
is a normalization constant. Using these definitions, we have
\begin{eqnarray}
F_0(y)=I_n(\upsilon),\quad\quad G_0(y)=I_n(\upsilon).\label{eq56}
\end{eqnarray}
Then, after finding the functions $F_0(y)$ and $G_0(y)$, we are able to proceed with the next perturbative approach.

Returning to Eq. (\ref{eq54}A3), we should solve the equation
\begin{eqnarray}
\mathbb{L}_1\left\{F_1\right\}=-2p_z\sqrt{2neB}I_{n-1}(\upsilon).\label{eq58}
\end{eqnarray} 
This is a non-homogeneous Hermite equation, whose solution will be $F_1=I_n(\upsilon)+f_p$. In other words, the solution is composed of the sum of the homogeneous equation given by (\ref{eq56}) and a particular function $f_p$ that needs to be determined. This function can be obtained using the method of undetermined coefficients. After applying this method, the complete solution for (\ref{eq54}A3) is
\begin{eqnarray}
F_1(\upsilon)=I_n(\upsilon)-p_z\left(\frac{2n}{eB}\right)^{\frac{1}{2}}I_{n-1}(\upsilon).
\end{eqnarray}
In a similar manner, for condition (\ref{eq54}A4), we will have
\begin{eqnarray}
G_1(\upsilon)=I_n(\upsilon)-p_z\left[\frac{2(n+1)}{eB}\right]^{\frac{1}{2}}I_{n+1}(\upsilon).
\end{eqnarray}
Notice that, while the solutions for $F$ and $G$ are very similar, the major difference between them lies in the dispersion relation.

Using these results, Eqs. (\ref{eq59}) and (\ref{eq60}) become
\begin{eqnarray}
f_1(y)&=&I_n(\upsilon)+d_{33}\left[I_n(\upsilon)-p_z\left(\frac{2n}{eB}\right)^{\frac{1}{2}}I_{n-1}(\upsilon)\right];\nonumber\\ f_2(y)&=&I_n(\upsilon)+d_{33}\left[I_n(\upsilon)-p_z\left(\frac{2(n+1)}{eB}\right)^{\frac{1}{2}}I_{n+1}(\upsilon)\right].
\end{eqnarray}
Then, the left-handed solution
$\psi_L=\begin{pmatrix}
\phi_1(y)\\\phi_2(y)
\end{pmatrix}$, with $\phi_k=e^{i\vec{p}\cdot\vec{r}_{\cancel{y}}}f_k(y)$, is obtained. As $\phi_1$ and $\phi_2$ have different energy eigenvalues and represent different states, it is necessary to split them in order to form the left-handed spinor and, consequently, the right-handed ones.

\subsubsection{Right-handed solutions}

In this subsection, let's determine the right-handed solution. Then,  Eq. (\ref{eq02}b) is rewritten as
\begin{eqnarray}
m\psi_R=E\psi_L+\sigma^1\left(i\partial^1+eA^1\right)\psi_L+\sigma^2\left(i\partial^2\right)\psi_L+\left(1-d_{33}\right)\sigma^3\left(i\partial^3\right)\psi_L. \label{eq61}
\end{eqnarray}

As mentioned before, it is necessary to split up the components of $\psi_L$. When we choose $\psi_L=\phi_1$, the solutions for the right-handed spinor will be $\psi_R=\begin{pmatrix}
\chi_1^{(1)}\\\chi_2^{(1)}
\end{pmatrix}$. Therefore, by calculating the terms of Eq. (\ref{eq61}) separately and suppressing the exponential of plane waves $e^{i\vec{p}\cdot\vec{r}_{\cancel{y}}}$, we have
\begin{eqnarray}
\chi_1^{(1)}=\frac{1}{m}\left[\left(1+d_{33}\right)\left(E+p_z\right)-d_{33}p_z\right]I_n-\frac{d_{33}p_z(E+p_z)}{m}\left(\frac{2n}{eB}\right)^{\frac{1}{2}}I_{n-1}
\end{eqnarray}
and
\begin{eqnarray}
\chi_2^{(1)}=-\frac{\sqrt{2(n+1)eB}}{m}(1+d_{33})I_{n+1}+\frac{d_{33}p_z}{m}\sqrt{4n^2}I_{n}.
\end{eqnarray}

On the other hand, when we choose the solution $\psi_L=\phi_2$, the right-handed solution becomes $\psi_R=\begin{pmatrix}
\chi_1^{(2)}\\\chi_2^{(2)}
\end{pmatrix}$, with
\begin{eqnarray}
\chi_1^{(2)}=-\frac{\sqrt{2neB}}{m}(1+d_{33})I_{n-1}+\frac{d_{33}p_z}{m}\sqrt{4(n+1)^2}I_{n}
\end{eqnarray}
and
\begin{eqnarray}
\chi_2^{(2)}=\frac{1}{m}\left[\left(1+d_{33}\right)\left(E-p_z\right)+d_{33}p_z\right]I_n-\frac{d_{33}p_z(E-p_z)}{m}\left(\frac{2(n+1)}{eB}\right)^{\frac{1}{2}}I_{n+1}.
\end{eqnarray}

Finally, we can write the particle solutions to the Dirac Equation
\begin{eqnarray}
\psi=e^{-iEt+ip_xx+ip_zz}\begin{pmatrix}
I_n+d_{33}\left[I_n-p_z\left(\frac{2n}{eB}\right)^{\frac{1}{2}}I_{n-1}\right]\\
0\\
\frac{1}{m}\left[\left(1+d_{33}\right)\left(E+p_z\right)-d_{33}p_z\right]I_n-\frac{d_{33}p_z(E+p_z)}{m}\left(\frac{2n}{eB}\right)^{\frac{1}{2}}I_{n-1}\\
-\frac{\sqrt{2(n+1)eB}}{m}(1+d_{33})I_{n+1}+\frac{d_{33}p_z}{m}\sqrt{4n^2}I_{n}
\end{pmatrix}\label{eq64}
\end{eqnarray}
whose energy is $E=\sqrt{m^2+p_z^2+2(n+1)eB}+p_zd_{33}$ and
\begin{eqnarray}
\psi=e^{-iEt+ip_xx+ip_zz}\begin{pmatrix}
0\\
I_n+d_{33}\left[I_n-p_z\left(\frac{2(n+1)}{eB}\right)^{\frac{1}{2}}I_{n+1}\right]\\
-\frac{\sqrt{2neB}}{m}(1+d_{33})I_{n-1}+\frac{d_{33}p_z}{m}\sqrt{4(n+1)^2}I_{n}\\
\frac{1}{m}\left[\left(1+d_{33}\right)\left(E-p_z\right)+d_{33}p_z\right]I_n-\frac{d_{33}p_z(E-p_z)}{m}\left(\frac{2(n+1)}{eB}\right)^{\frac{1}{2}}I_{n+1}
\end{pmatrix}\label{eq65}
\end{eqnarray}
whose energy is $E=\sqrt{m^2+p_z^2+2neB}-p_zd_{33}$.

Analyzing in more detail, we can observe that for $n=0$, we will have contributions of $\sqrt{m^2+p_z^2+2eB}$ and $\sqrt{m^2+p_z^2}$ for the energies of the solutions (\ref{eq64}) and (\ref{eq65}), respectively. However, for $n=1$, these contributions become $\sqrt{m^2+p_z^2+4eB}$ and $\sqrt{m^2+p_z^2+2eB}$, where in this case the latter has the same value as the former for $n=0$. Therefore, as both contributions are related to $I_n$, we can standardize this problem by defining $\varepsilon=\sqrt{m^2+p_z^2+2neB}$ and assigning the first solution to $I_n\to I_{n-1}$. In other words, we rewrite the solutions given by Eqs. (\ref{eq64}) and (\ref{eq65}) as
\begin{eqnarray}
\psi=e^{-ip\cdot x_{\cancel{y}}}\begin{pmatrix}
I_{n-1}+d_{33}\left[I_{n-1}-p_z\left(\frac{2(n-1)}{eB}\right)^{\frac{1}{2}}I_{n-2}\right]\\
0\\
\frac{1}{m}\left[\left(1+d_{33}\right)\left(E+p_z\right)-d_{33}p_z\right]I_{n-1}-\frac{d_{33}p_z(E+p_z)}{m}\left(\frac{2(n-1)}{eB}\right)^{\frac{1}{2}}I_{n-2}\\
-\frac{\sqrt{2neB}}{m}(1+d_{33})I_{n}+\frac{d_{33}p_z}{m}2(n-1)I_{n-1}
\end{pmatrix}\label{eq72}
\end{eqnarray}
for the energy $E=\varepsilon+p_zd_{33}$ and
\begin{eqnarray}
\psi=e^{-ip\cdot x_{\cancel{y}}}\begin{pmatrix}
0\\
I_n+d_{33}\left[I_n-p_z\left(\frac{2(n+1)}{eB}\right)^{\frac{1}{2}}I_{n+1}\right]\\
-\frac{\sqrt{2neB}}{m}(1+d_{33})I_{n-1}+\frac{d_{33}p_z}{m}\sqrt{4(n+1)^2}I_{n}\\
\frac{1}{m}\left[\left(1+d_{33}\right)\left(E-p_z\right)+d_{33}p_z\right]I_n-\frac{d_{33}p_z(E-p_z)}{m}\left(\frac{2(n+1)}{eB}\right)^{\frac{1}{2}}I_{n+1}
\end{pmatrix}\label{eq73}
\end{eqnarray}
for the energy $E=\varepsilon-p_zd_{33}$. Where $I_{-1}=I_{-2}=0$.

Considering negative-energy particles and proceeding in a very similar way, solutions for the anti-particle spinors are obtained as
\begin{eqnarray}
\psi=e^{ip\cdot r_{\cancel{y}}}\begin{pmatrix}
\frac{1}{m}\left[(1+d_{33})(p_z-E)-d_{33}p_z\right]I_{n-1}-\frac{d_{33}p_z(p_z-E)}{m}\left[\frac{2(n-1)}{eB}\right]^{\frac{1}{2}}I_{n-2}\\
\frac{\sqrt{2neB}}{m}(1+d_{33})I_{n}-\frac{d_{33}p_z}{m}[2(n-1)]I_{n-1}\\
I_{n-1}(\tilde{\upsilon})+d_{33}\left[I_{n-1}(\tilde{\upsilon})-p_z\left(\frac{2(n-1)}{eB}\right)^{\frac{1}{2}}I_{n-2}(\tilde{\upsilon})\right]\\
0
\end{pmatrix}\label{eq74}
\end{eqnarray}
for the energy $E=\varepsilon+p_zd_{33}$ and
\begin{eqnarray}
\psi=e^{ip\cdot r_{\cancel{y}}}\begin{pmatrix}
\frac{\sqrt{2neB}}{m}(1+d_{33})I_{n-1}-\frac{d_{33}p_z}{m}2(n+1)I_n\\
-\frac{1}{m}\left[(1+d_{33})(E+p_z)+d_{33}p_z\right]I_n+\frac{d_{33}p_z(E+p_z)}{m}\left[\frac{2(n+1)}{eB}\right]^{\frac{1}{2}}I_{n+1}\\
0\\
I_n(\tilde{\upsilon})+d_{33}\left[I_n(\tilde{\upsilon})-p_z\left(\frac{2(n+1)}{eB}\right)^{\frac{1}{2}}I_{n+1}(\tilde{\upsilon})\right]
\end{pmatrix}\label{eq75}
\end{eqnarray}
for the energy $E=\varepsilon-p_zd_{33}$.

In the next subsection, the solutions for particles and antiparticles are analyzed for the limit of a strong external magnetic field.

\subsection{Strong field limits}\label{secstrongerfields}

When we consider external fields with higher magnitudes, the range of possible values of $n$ decreases rapidly until reaching the limit of $n=0$ as the only possible level \cite{magfieldthesis, danielmag}. For the following analysis, let's set $n=0$ for our problem and begin examining the spinor expressions. In this manner, Eqs. (\ref{eq73}) and (\ref{eq75}) can be rewritten as
\begin{eqnarray}
\psi\equiv N_ue^{-ip\cdot r_{\cancel{y}}}u_2=N_ue^{ip\cdot r_{\cancel{y}}}\begin{pmatrix}
0\\(1+d_{33})I_0-d_{33}p_z\alpha I_1\\2\frac{d_{33}p_z}{m}I_0\\\frac{1}{m}\left[(1+d_{33})(E-p_z)+d_{33}p_z\right]I_0-\frac{d_{33}p_z(E-p_z)}{m}\alpha I_1
\end{pmatrix}\label{eq85}
\end{eqnarray}
and
\begin{eqnarray}
\psi\equiv N_ve^{ip\cdot r_{\cancel{y}}}v_2=N_ve^{ip\cdot r_{\cancel{y}}}\begin{pmatrix}
-2\frac{d_{33}p_z}{m}I_0\\-\frac{1}{m}\left[(1+d_{33})(E+p_z)+d_{33}p_z\right]I_0+\frac{d_{33}p_z(E+p_z)}{m}\alpha I_1\\0\\(1+d_{33})I_0-d_{33}p_z\alpha I_1
\end{pmatrix}.\label{eq86}
\end{eqnarray}
With $N_u$ and $N_v$ being normalization constants and $\alpha=\sqrt{\frac{2}{eB}}$. Both Eqs. (\ref{eq72}) and (\ref{eq74}) are equal to zero, indicating that the only possible solutions for particles and antiparticles are those given by Eqs. (\ref{eq85}) and (\ref{eq86}), respectively. With this established, we can now proceed to calculate a scattering process in QED under an external field and account for Lorentz violation effects.

\section{The scattering}\label{secscattering}

In this section, the main objective is to calculate the cross section for the  $e^+ e^-\rightarrow l^+ l^-$ scattering and analyze the corrections to this quantity due to Lorentz violation, EDM, and the presence of an external magnetic field. Returning to the initial Lagrangian, Eq. (\ref{eq76}),
\begin{eqnarray}
\mathcal{L}=\bar{\psi}i\left[\left(\partial_\mu+d_{\mu\nu}\gamma^5\partial^\nu\right)-ie\left(A_\mu+d_{\mu\nu}\gamma^5 A^\nu\right)\right]\gamma^\mu\psi-m\bar{\psi}\psi,
\end{eqnarray}
we can write the interaction Lagrangian as
\begin{eqnarray}
\mathcal{L}_{\text{int}}=e\bar{\psi}\left(A_{\mu}+d_{\mu\nu}\gamma^5A^{\nu}\right)\gamma^\mu\psi.\label{eq77}
\end{eqnarray}

From Eq. (\ref{eq77}) we can extract an interacting current $j^{\mu}$ by analogy with the coupling term $j_\mu A^{\mu}$. However, to take lepton EDM into account in the problem, it is necessary to return at the form factors given by Eq. (\ref{eq78}) and pay a special attention to that which represents this contribution, i. e., the term $\frac{F_3(q^2)}{2m}\sigma^{\mu\nu}q^{\nu}\gamma^5$. By momentum conservation, it is simple to see that the zero recoil of momentum transfer condition $q^2=0$ is already satisfied. Therefore, it is necessary to make a change in the coupling current, and the interaction lagrangian (\ref{eq77}) becomes
\begin{eqnarray}
\mathcal{L}_{\text{int}}&\equiv& e\bar{\psi}\left[\left(A_{\mu}+d_{\mu\nu}\gamma^5 A^{\nu}\right)\gamma^\mu+\frac{F_3(0)}{2m}\sigma_{\mu\nu}q^\nu A^\mu\gamma^5\right]\psi\nonumber\\
&=&e\bar{\psi}\left(A_{\mu}\gamma^\mu+d_{\mu\nu}\gamma^5 A^{\nu}\gamma^\mu-\varrho_l\sigma_{\mu\nu}q^\nu A^\mu\gamma^5\right)\psi,
\end{eqnarray}
where $\varrho_l$ is the lepton EDM, as shown in Eq. (\ref{eq79}). 

\begin{figure}[ht]
  \centering
  \subfigure[]{\includegraphics[width=0.32\textwidth]{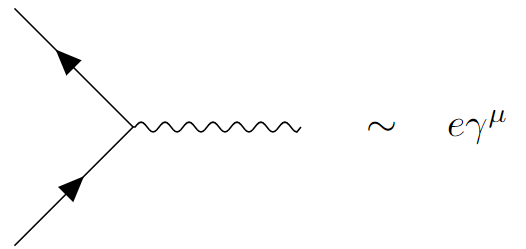}}
  \hfill
  \subfigure[]{\includegraphics[width=0.32\textwidth]{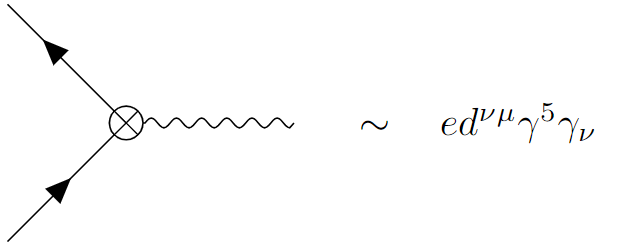}}
  \hfill
  \subfigure[]{\includegraphics[width=0.32\textwidth]{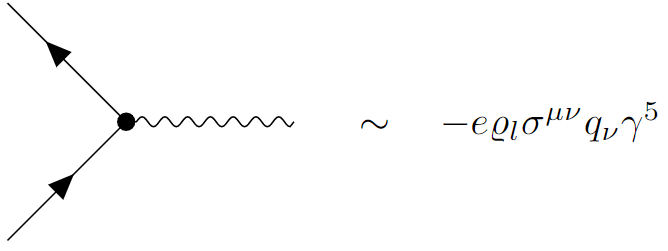}}
  \caption{Three types of interaction vertices are represented: (a) the usual QED, (b) the LV-contribution, and (c) the lepton EDM consideration.\label{fig5}}
\end{figure}

Notice that we are dealing with an extension of the usual QED beyond the Standard Model. That is, a theory involving electrons, positrons, and photons subject to a very strong external field and the effects of a nature that violates Lorentz symmetry through the coefficient $d_{33}$. In this new theory, there are three types of vertices, as mentioned before. The diagrams of these vertices are represented in Figure \ref{fig5}.

The classical external field is present in the problem through the leptonic field operator. In a more general way it is given by \cite{magfield}
\begin{eqnarray}
\psi=\int\frac{dp_xdp_z}{(2\pi)^2}\sum_{n,s}N_n\left[a_su_se^{-ip\cdot r_{\cancel{y}}}+b_s^\dagger v_s e^{ip\cdot r_{\cancel{y}}}\right].
\end{eqnarray}
In the present case, it reduces to
\begin{eqnarray}
\psi=\int\frac{dp_z}{2\pi}\left[N_u(p_z)a(p_z)u(y,p_z)e^{-ip\cdot r_{\cancel{y}}}+N_v(p_z)b^\dagger(p_z) v(y,p_z) e^{ip\cdot r_{\cancel{y}}}\right].
\end{eqnarray}

\begin{figure}[ht]
\includegraphics[scale=0.5]{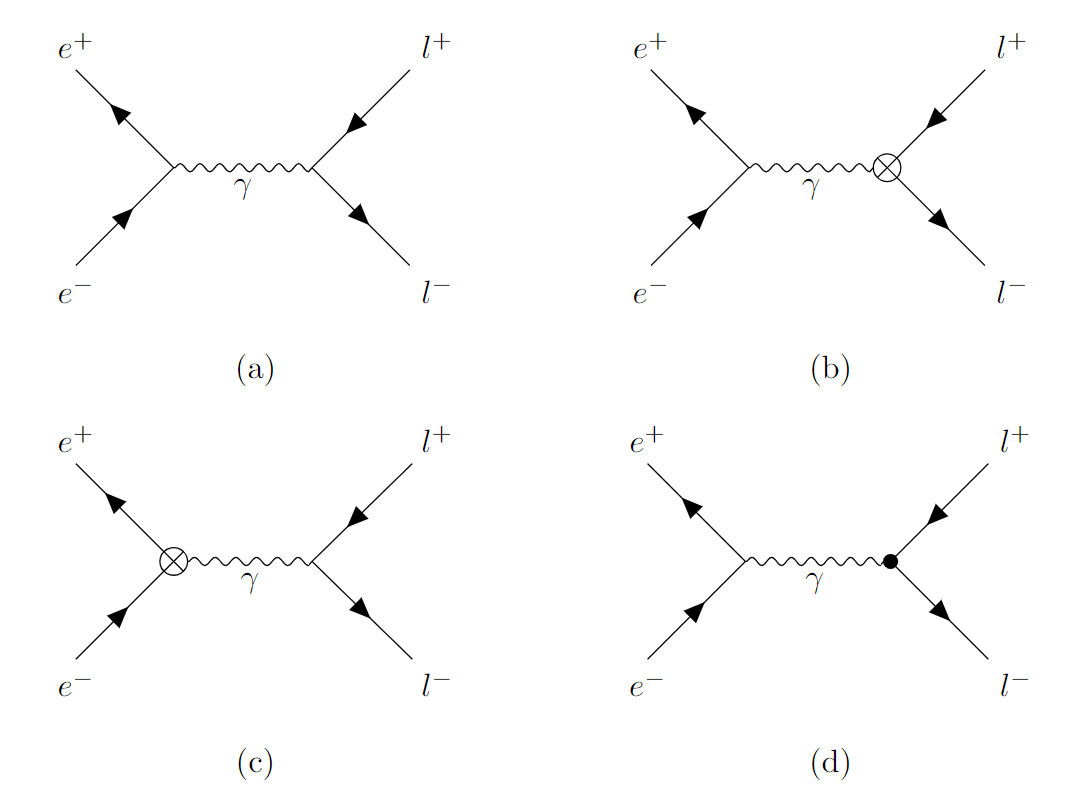}
\caption{Representation of scattering processes: (a) Represents the usual QED scattering. (b) and (c) display LV-processes. (d) Represents the contribution from the final lepton EDM.}
\label{fig1}
\end{figure}

Taking a scattering process of the type $e^{+}e^{-}\to l^{+}l^{-}$ when only the final leptons contribution of the EDM is considered, the diagrams that describe this process are given in the Figure \ref{fig1}.

It is important to note that, typically, for an interaction Lagrangian with three terms, the scattering would require nine diagram contributions: the usual QED, three LV-corrections (two of first order and one of second), three EDM-corrections (one for each vertex separately and one for both vertices simultaneously), and finally, two LV-EDM contributions. However, since the EDM depends directly on the  $d_{\mu\nu}$ term and we are considering contributions only up to the lowest possible order, the number of diagrams drops to five. If we consider only the EDM of the final lepton, we need to study only the graphs presented in Figure \ref{fig1}. The reason for discarding the electronic contribution of the dipole moment is that, as mentioned before, this quantity is much lower compared to the heavier leptons, especially at high energies.

The second order element of the scattering operator is given as
{\small
\begin{eqnarray}
\hat{S}^{(2)}&=&\frac{(-i)^2}{2}\int d^4r_1d^4r_2 \mathcal{T}\left[\mathcal{L}_{\text{int}}(r_1)\mathcal{L}_{\text{int}}(r_2)\right]\nonumber\\
&=&-\frac{e^2}{2}\int d^4r_1d^4r_2 \mathcal{T}\left[\bar{\psi}_1A_{1\beta}\gamma^\beta\psi_1\bar{\psi}_2A_{2\beta}\gamma^\beta\psi_2\right]-\frac{e^2}{2}\int d^4r_1d^4r_2\mathcal{T}\left[\bar{\psi}_1A_{1\beta}\gamma^\beta\psi_1\bar{\psi}_2 d_{\mu\nu}\gamma^5A_2^\nu\gamma^\mu \psi_2\right]
\nonumber\\
&&-\frac{e^2}{2}\int d^4r_1d^4r_2 \mathcal{T}\left[\bar{\psi}_1d_{\mu\nu}\gamma^5A_1^\nu\gamma^\mu\psi_1\bar{\psi}_2 A_{2\beta}\gamma^\beta \psi_2\right]+\frac{e^2}{2}\int d^4r_1d^4r_2 \mathcal{T}\left[\bar{\psi}_1A_{1\beta}\gamma^\beta\psi_1\bar{\psi}_2\varrho_l\sigma_{\mu\nu}q^\nu A_2^\mu\gamma^5\psi_2\right]\nonumber\\
&\equiv & \mathcal{M}_a+\mathcal{M}_b+\mathcal{M}_c+\mathcal{M}_d
\end{eqnarray}
}
with $\mathcal{T}$ being the time order operator.

Assuming that the initial and final states are given by 
\begin{eqnarray}
\ket{i}=a^{\dagger}(p_z)b^\dagger(k_z)\ket{0};\quad\quad \ket{f}=a^\dagger(p^\prime_z)b^\dagger(k^\prime_z)\ket{0},
\end{eqnarray}
the scattering amplitude $\mathcal{M}_{fi}=\bra{f}\hat{S}^{(2)}\ket{i}$ will have the contribution $\mathcal{M}_{fi}=\mathcal{M}_{a}+\mathcal{M}_{b}+\mathcal{M}_{c}+\mathcal{M}_{d}$. The first contribution reads
\begin{eqnarray}
\mathcal{M}_a&=&-\frac{e^2}{2}\int d^4r_1d^4r_2 \bra{f}\mathcal{T}\left[\bar{\psi}_1A_{1\beta}\gamma^\beta\psi_1\bar{\psi}_2A_{2\beta}\gamma^\beta\psi_2\right]\ket{i}\nonumber\\
&=&-ie^2\int dy\left[\bar{v}(k_z)\gamma_\mu u(p_z)\right]\frac{g^{\mu\nu}}{s}\left[\bar{u}(p^\prime_z)\gamma_\nu v(k^\prime_z)\right].\label{61}
\end{eqnarray}
Here the photon propagator definition, i.e.,
\begin{eqnarray}
\bra{0}\mathcal{T}\left[A_1^\nu A_2^\beta\right]\ket{0}=i\int\frac{d^4\kappa}{(2\pi)^4}e^{-i(r_2-r_1)\kappa}\frac{g^{\nu\beta}}{\kappa^2},
\end{eqnarray}
and
\begin{eqnarray}
\left\{a(p),a^\dagger(k)\right\}=\frac{(2\pi)}{N_u}\delta(p_z-k_z),\quad\quad \left\{b(p),b^\dagger(k)\right\}=\frac{(2\pi)}{N_v}\delta(p_z-k_z)
\end{eqnarray}
have been used. The delta function $(2\pi)^3\delta_{\cancel{y}}^4(p_2+k_2-p_1-k_1)$ was suppressed. It represents the momentum conservation. 

In a similar way, the other scattering amplitude contributions are given as
\begin{eqnarray}
\mathcal{M}_b&=&-ie^2\int dy\left[\bar{v}(k_z)\gamma_\beta u(p_z)\right]\frac{g^{\beta\nu}}{s}\left[\bar{u}(p^\prime_z)d_{\mu\nu}\gamma^5\gamma^\mu v(k^\prime_z)\right],\\
\mathcal{M}_c&=&-ie^2\int dy\left[\bar{v}(k_z)d_{\mu\nu}\gamma^5\gamma^\mu u(p_z)\right]\frac{g^{\nu\beta}}{s}\left[\bar{u}(p^\prime_z)\gamma_{\beta}v(k^\prime_z)\right],\\
\mathcal{M}_d&=&ie^2\int dy\left[\bar{v}(k_z)\gamma_\beta u(p_z)\right]\frac{g^{\beta\mu}}{s}\left[\bar{u}(p^\prime_z)\varrho_l\sigma_{\mu\nu}q^\nu\gamma^5 v(k^\prime_z)\right].
\end{eqnarray}
Furthermore, we consider a frame in which the momenta become
\begin{eqnarray}
p=(E_i(\vec{p}_i),\vec{p}_i);\quad\quad k=(E_i(-\vec{p}_i),-\vec{p}_i).\quad\quad p^{\prime}=(E_f(\vec{p}_f),\vec{p}_f);\quad\quad k^\prime=(E_f(-\vec{p}_f),-\vec{p}_f).
\end{eqnarray}
Then, the contribution given in Eq. (\ref{61}) is written as
\begin{eqnarray}
\mathcal{M}_a&=&4ie^2\frac{1+2\left[d_{33}^{(e)}+d_{33}^{(l)}\right]}{s}\frac{\varepsilon^2-\varepsilon(p_i+p_f)+p_ip_f}{m_em_l}\int I_0^4dy\nonumber\\
&=&4ie^2\sqrt{\frac{eB}{2\pi}}\frac{1+2\left[d_{33}^{(e)}+d_{33}^{(l)}\right]}{s}\frac{\varepsilon^2-\varepsilon(p_i+p_f)+p_ip_f}{m_em_l},
\end{eqnarray}
where, in the strong field limit, it has been used
\begin{eqnarray}
\int I_0^4dy&=&\int\left[\left(\frac{\sqrt{eB}}{\sqrt{\pi}}\right)^{\frac{1}{2}}e^{\frac{-\upsilon^2}{2}}H_0(\upsilon)\right]^{4}dy=\sqrt{\frac{eB}{2\pi}}.
\end{eqnarray}

For the other diagrams we obtain, up to lowest order on LV-coefficient,
\begin{eqnarray}
\mathcal{M}_b&=&4i\sqrt{\frac{eB}{2\pi}}\frac{e^2d_{33}^{(l)^3}}{s}\frac{p_ip_f}{m_em_l},\\
\mathcal{M}_c&=&4i\sqrt{\frac{eB}{2\pi}}\frac{e^2d_{33}^{(e)^3}}{s}\frac{p_ip_f}{m_em_l},\\
\mathcal{M}_d&=&4\sqrt{\frac{eB}{2\pi}}\biggl\{1+2\left[d_{33}^{(e)}+d_{33}^{(l)}\right]\biggr\}\frac{e^2\varrho_l\varepsilon}{s}\frac{(\varepsilon-p_i)[\varepsilon^2+p_f(p_f-2\varepsilon)-m_l^2]}{m_em_l^2}.
\end{eqnarray}

Then, the total scattering amplitude becomes
\begin{eqnarray}
\mathcal{M}_{fi}&=&i\frac{4e^2}{sm_em_l}\sqrt{\frac{eB}{2\pi}}\left\{\left[1+2\left(d_{33}^{(e)}+d_{33}^{(l)}\right)\right]\left[\varepsilon^2-\varepsilon(p_i+p_f)+p_ip_f\right]+\left[d_{33}^{(e)^3}+d_{33}^{(l)^3}\right]p_ip_f\right.\nonumber\\
&-&\left.i\frac{\varrho_l\varepsilon}{m_l}\left[1+2\left(d_{33}^{(e)}+d_{33}^{(l)}\right)\right](\varepsilon-p_i)\left[\varepsilon^2+p_f(p_f-2\varepsilon)-m_l^2\right]\right\}.
\end{eqnarray}
To calculate the cross section, the most important quantity is the squared scattering amplitude. Considering up to third order in Lorentz violation coefficients, it is given as
\begin{eqnarray}
|\mathcal{M}_{fi}|^2&=&\frac{16e^4}{s^2m_e^2m_l^2}\frac{eB}{2\pi}\biggl\{\biggl[1+2\bigg(d_{33}^{(e)}+d_{33}^{(l)}\bigg)\biggr]^2(\varepsilon-p_f)^2(\varepsilon-p_i)^2+2\biggl(d_{33}^{(e)^3}+d_{33}^{(l)^3}\biggr)(\varepsilon-p_f)(\varepsilon-p_i)p_ip_f\nonumber\\&+&\bigg[1+4\biggl(d_{33}^{(e)}+d_{33}^{(l)}\biggr)\bigg]\frac{|\varrho_l |^2\varepsilon^2}{m_l^2}(\varepsilon-p_i)^2\bigg[(\varepsilon-p_f)^2-m_l^2\bigg]\biggr\},\label{74}
\end{eqnarray}
where was used $E_{i/f}=\varepsilon_{i/f}-d^{(i/f)}_{33}p_{i/f}$, but by the momentum conservation $p_1+k_1=p_2+k_2\to\varepsilon_1=\varepsilon_2\equiv\varepsilon$.

For a given scattering process, the cross section is defined by the relation \cite{danielmag}
\begin{eqnarray}
\sigma&=&\frac{|\mathcal{M}_{fi}|^2}{8p_fs\sqrt{s-4m_e^2}}.
\end{eqnarray}
Therefore, using Eq. (\ref{74}) the cross section becomes
{\small
\begin{eqnarray}
\sigma&=&\frac{8\alpha^2(\pi e B)}{s^3m_e^2m_l^2\sqrt{\varepsilon^2-m_e^2}\sqrt{\varepsilon^2-m_l^2}}\biggl\{\biggl[1+2\bigg(d_{33}^{(e)}+d_{33}^{(l)}\bigg)\biggr]^2(\varepsilon-p_f)^2(\varepsilon-p_i)^2\nonumber\\&+&2\biggl(d_{33}^{(e)^3}+d_{33}^{(l)^3}\biggr)(\varepsilon-p_f)(\varepsilon-p_i)p_ip_f+\bigg[1+4\biggl(d_{33}^{(e)}+d_{33}^{(l)}\biggr)\bigg]\frac{|\varrho_l |^2\varepsilon^2}{m_l^2}(\varepsilon-p_i)^2\bigg[(\varepsilon-p_f)^2-m_l^2\bigg]\biggr\}
,\label{eq83}
\end{eqnarray}
} where we have used $s=4\varepsilon^2$ and $\alpha=e^2/4\pi$. 

This result shows the corrected cross section due to the external magnetic field, EDM, and Lorentz violation. Based on this result, some points should be discussed.

With Eqs. (\ref{eq79}) and (\ref{eq82}), we can express the EDM enhancement as
\begin{eqnarray}
|\varrho_l| = \frac{275\alpha}{36\pi}\frac{d_{33}^{(l)}p_f^2}{m_l^3},\label{eq88}
\end{eqnarray}
where the subscript $l$ specifies the type of lepton that is considered.
Restoring the correct units, we will have for the electron
\begin{eqnarray}
|\varrho_{e}| = 6.85\times10^{-13}\frac{d_{33}^{(e)}p_f^2}{m_e^2}e.\text{cm}.
\end{eqnarray}
Similarly, for the muon and tau leptons we have, respectively,
\begin{eqnarray}
|\varrho_{\mu}| = 3.31\times10^{-15}\frac{d_{33}^{(\mu)}p_f^2}{m_\mu^2}e.\text{cm},\quad\quad |\varrho_\tau| = 1.96\times10^{-16}\frac{d_{33}^{(\tau)}p_f^2}{m_{\tau}^2}e.\text{cm}.\label{eq87}
\end{eqnarray}

According to measurements \cite{edm03}, we have an upper limit for the $\mu$EDM of $1.8\times10^{-19}e.\text{cm}$ at a confidence level of $95\%$, at an energy around $3\, \text{GeV}$. Direct substitution into Eq. (\ref{eq87}) suggests $d_{33} < 6.76 \times 10^{-8}$. Similarly, for the $\tau$EDM of $4.94\times10^{-16}e.\text{cm}$ at a confidence level of $95\%$ and energy around $5\, \text{GeV}$ \cite{edm05}, we would have $d_{33} < 0.36$. However, it's worth noting that these estimates are based on limits synthesized by \cite{tableviolation}, and thus these orders of magnitude represent very rough estimates for the LV-coefficient. 

However, for those limits, in terms of energy, we have approximately $\mu$EDM $\sim 9.1\times10^{-6}\text{GeV}^{-1}$ and $\tau$EDM $\sim 2.5\times10^{-2}\text{GeV}^{-1}$. Furthermore, it's important to note from Eq. (\ref{eq83}) that the EDM contribution to the cross section appears as a factor $(|\varrho_l|\varepsilon)^2$. Therefore, for example, considering a reaction where the final products are tau particles with energy $\varepsilon=1 \, \text{TeV}$, the cross section will increase by a factor of $(|\varrho_\tau|\varepsilon)^2=625$, which represents a significant and impactful contribution.

\begin{figure}[ht]
  \centering

  \subfigure[\label{fig2}]{\includegraphics[width=0.49\textwidth]{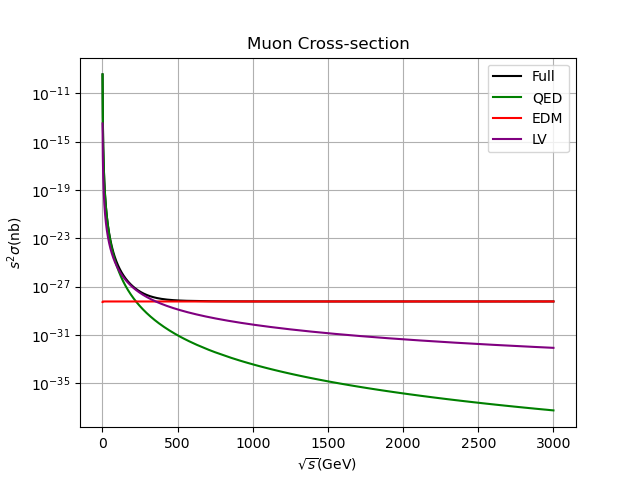}}
  \hfill
  \subfigure[\label{fig2b}]{\includegraphics[width=0.49\textwidth]{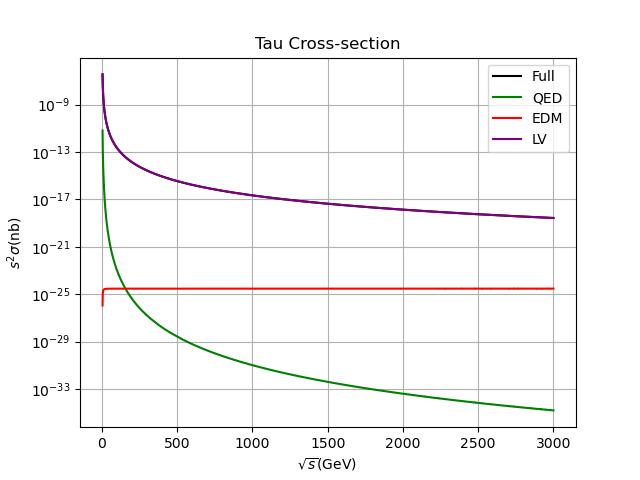}}

  \caption{The plots depict the full function of $s^2\sigma$ (in black), along with its constituent parts: QED contribution (in green), EDM contribution (in red), and Lorentz-violating contribution (in purple), separately for muon and tau final products..}\label{fig3}
\end{figure}

The enhancement in the EDM for leptons, as described in Eq. (\ref{eq88}), indicates that at very high energies, this quantity becomes a function of $\varepsilon^2$, implying that the EDM contribution to the cross-section starts to dominate, as illustrated in Figures \ref{fig2} and \ref{fig2b}. Consequently, we observe that the limits, at energies of the order of $\sim 1-100 \, \text{GeV}$, $\varrho_\mu < 1.8 \times 10^{-19} \, e \cdot \text{cm}$ and $\varrho_\tau < 4.94 \times 10^{-16} \, e \cdot \text{cm}$, become increasingly greater as the energy of the process increases. 

Consequently, if we consider a scattering process with muon particles, whose characteristic energies are very high, around $\varepsilon=\text{1TeV}$, such that the enhancement on the EDM, to explore the possibility, is of the order of $10^{-4}\text{GeV}^{-1}$, the correction in the total cross-section will be appro\-ximately $(|\varrho_\mu|\varepsilon)^2\sim1\%$. In other words, in this context, applying Eq. (\ref{eq88}) directly, we obtain the limit $d_{33}^{(\mu)}=6.65\times10^{-12}$. This assumption of the enhancement magnitude is lower than those proposed by \cite{edmweak} based on weak effects; consequently, the limit for the LV-coefficient is smaller than that obtained by \cite{edmexpression}. In the same context, for comparison, supposing that the enhancement of the $\tau$EDM is of the order of $\tau$EDM $\sim 2.4\times10^{-2}\text{GeV}^{-1}$ at $\varepsilon=1\text{TeV}$, we would obtain $d_{33}^{(\tau)}=7.6\times10^{-6}$.

Thus, we can summarize all the results collected here for all leptons in Table \ref{tab1}. This outcome justifies the decision to disregard the electron EDM in the scattering process; that is, both $d_{33}^{(e)}$ and $|\varrho_e|$ are very small, leading to a lower upper limit when compared with the other leptons.
\begin{table}[ht]
\begin{tabular}{|c|c|c|}
\hline 
Lepton flavor & LV-Coefficient\footnote{These LV limits were obtained by extrapolating to energies of $1$ 
TeV, for the muon and tau leptons.} & EDM top limit\footnote{These EDM limits were obtained for lower energies: around $3$ GeV for the muon and $5$ GeV for the tau.}\\\hline
Electron & $ <3\times10^{-15}$ \cite{tableviolation} & $<1.1\times10^{-29}e.\text{cm}$ \cite{electronedm}\\\hline
Muon & $ <6.65\times10^{-12}$  & $<1.8\times10^{-19}e.\text{cm}$ \cite{edm03}\\\hline
Tau & $ <7.6\times10^{-6}$  & $<4.94\times10^{-16}e.\text{cm}$ 
 \cite{edm05}\\\hline
\end{tabular}
\caption{The estimation of the upper limit for the LV coefficient and EDM values for all leptons.}\label{tab1}
\end{table}  

\begin{figure}[ht]
  \centering

  \subfigure[]{\includegraphics[width=0.49\textwidth]{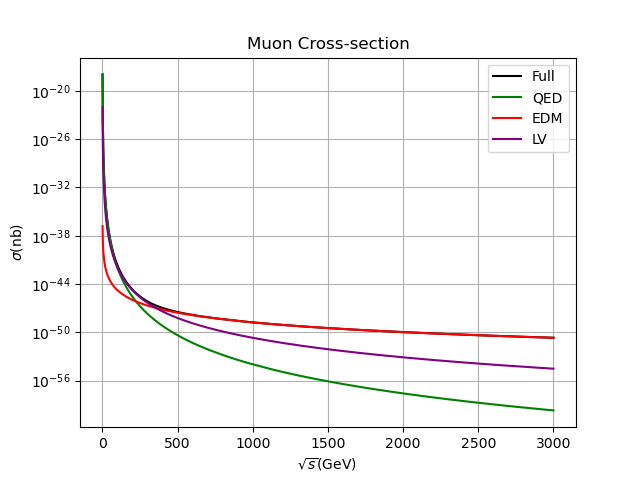}}
  \hfill
  \subfigure[]{\includegraphics[width=0.49\textwidth]{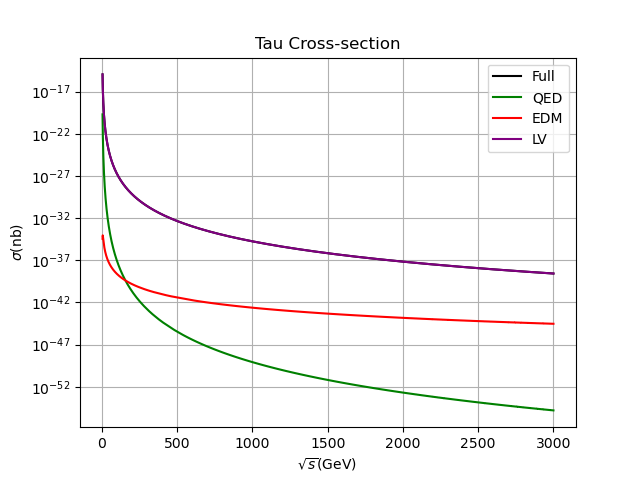}}

  \caption{The plots depict the cross-section $\sigma$ full function (in black), along with its constituent parts: QED contribution (in green), EDM contribution (in red), and Lorentz-violating contribution (in purple), separately for muon and tau final products.}\label{fig4}
\end{figure}

Figure \ref{fig3} illustrates the behavior of $s^2\sigma$ in terms of the center-of-mass energy. The additional dependence of the cross-section, as given in Eq. (\ref{eq83}), on $s^{-2}$ arises from the chiral representation of a Dirac theory in the presence of an external field. The formalism, where $\gamma^5$ is a diagonal matrix, introduces this additional dependence on the inverse Mandelstam variable \cite{chiralbook}. Additionally, the second dependence on $s^{-1}$ originates from the formalism involving an external field, as developed by \cite{magfieldthesis} and calculated by \cite{danielmag}. We restored the correct unit by multiplying Eq. (\ref{eq83}) by $m_e^2m_l^2$ for both cases, as illustrated in Figure \ref{fig4}. In all plots, a magnetic field strength of $eB\sim5.916\times10^{-5}\text{GeV}^2$ was used. This value corresponds to a transformation of natural units from a magnetic field magnitude of $B\sim10^{12}\text{T}$, which is typical in an astrophysical context \cite{field1, field2}. For all considerations, we took $e=1$.

This process was previously discussed by \cite{process}. However, in that case, the LV-coefficient was only taken into consideration in the composition of the lepton EDM, leading to one less vertex. Such an approach was also taken by \cite{magresult}, but without the presence of the $\varrho_l$ contribution. Additionally, since the spinors were not modified by the presence of the $d_{\mu\nu}$ tensor, the most convenient representation is the standard one, where $\gamma^0$ is a diagonal matrix. This explains the difference in results, despite the same functional behavior for the cross-section.

\section{Conclusions}\label{secconclusion}

The annihilation process of an electron-positron pair decaying into a heavier lepton-antilepton pair was examined. The formalism adopted began with a Lorentz-violating Dirac field in the presence of an external classical magnetic field oriented in the $z$-direction, using the chiral representation. The breakdown of Lorentz symmetries was introduced into the field via the term $d_{33}$ and an additional vertex in the scattering diagrams. The EDM of the final leptons, calculated from quantum corrections, is enhanced depending on the LV-coefficient and energy, manifesting as an additional vertex in the scattering process. The new spinor form is a function of energy, momenta, and a discrete index $n$ used for energy quantization, with a dependence on the spatial coordinate $y$ through a new function $I_n$. The limit of stronger fields is approached by setting $n=0$. Furthermore, the energy arises from a new dispersion relation that splits the energy of spin-up and spin-down solutions for particles and antiparticles by $\pm d_{33}p_z$.

Experimental data on the upper limits of EDM values, along with the expression for EDM enhancement, are utilized to establish upper limits of the LV-parameter $d_{33}$ for both the muonic and tauonic cases. These values are then compared with an extensive list of top limits on LV-parameters. While there are numerous results available for the electronic sector, data for the heavier leptons are limited. Therefore, this study provides an estimation of the $d_{33}$ value for all leptons.

Furthermore, the cross-section was calculated and plotted for $e^{+}e^{-}\to\mu^{+}\mu^{-}$ and $e^{+}e^{-}\to\tau^{+}\tau^{-}$ scatterings. In our calculations, at high energies around $\sqrt{s}\sim1\text{TeV}$, the cross-section $\sigma$ is predominantly influenced by the EDM contribution for the muon and by the LV contribution for the tau. In both cases, these contributions lead to significant corrections to the total cross-section, deviating from the expectations of conventional QED theory in the absence of these factors. Moreover, the presence of an external magnetic field, a natural component in astrophysical processes of this nature, serves to enhance the cross-section values and accentuate these additional effects compared to QED processes with $\Vec{B}=0$.

\section*{Acknowledgments}

This work by A. F. S. is partially supported by National Council for Scientific and Technological
Development - CNPq project No. 312406/2023-1. D. S. C. thanks CAPES for financial support.

\section*{Data Availability Statement}

%No data are available because of the nature of the research. This publication is theoretical work that does not require supporting research data.
No Data associated in the manuscript.
%%%%%%%%%%%%%%%%%%%%%%%%%%%%%%%%%%%%%%%%%%%%%%%%%%%%%%%%%%%%%%%%%%%%%%%%%%%%%%%%%%%%%%%%%%%%%%%%%%%%%%%%%%%%%%%%%

\global\long\def\link#1#2{\href{http://eudml.org/#1}{#2}}
 \global\long\def\doi#1#2{\href{http://dx.doi.org/#1}{#2}}
 \global\long\def\arXiv#1#2{\href{http://arxiv.org/abs/#1}{arXiv:#1 [#2]}}
 \global\long\def\arXivOld#1{\href{http://arxiv.org/abs/#1}{arXiv:#1}}

%%%%%%%%%%%%%%%%%%%%%%%%%%%%%%%%%%%%%%%%%%%%%%%%%%%%%%%%%%%%%%%%%%%%%%%%%%%%%%%%%%%%%%%%%%%%%%%%%%%%%%%%%%%


\begin{thebibliography}{99}

\bibitem{colladayredef} D. Colladay and V. A. Kosteleck\'y, ``Cross sections and lorentz violation,'' \doi{10.1016/S0370-2693(01)00649-9}{{Phys. Lett. B} {\bf 511}, 209 (2001)}.

\bibitem{kostelecky1} D. Colladay and V. A. Kosteleck\'y, ``Lorentz-violationg extension of the standard model,'' \doi{10.1103/PhysRevD.58.116002}{{Phys. Rev. D} {\bf 58}, 116002 (1998)}.

\bibitem{kostelecky2} D. Colladay and V. A. Kosteleck\'y, ``CPT violation and the standard model,'' \doi{10.1103/PhysRevD.55.6760}{{Phys. Rev. D} {\bf 55}, 6760 (1997)}.

\bibitem{ale1} A. F. Santos and F. C. Khanna. "Lorentz violation in Bhabha scattering at finite temperature." \doi{10.1103/PhysRevD.95.125012}{{Phys. Rev. D} {\bf 95}, 125012 (2017)}.

\bibitem{daniel1} D. S. Cabral, A. F. Santos and F. C. Khanna, ``Violation of Lorentz symmetries and thermal effects in Compton scattering,'' \doi{10.1140/epjp/s13360-023-03707-w}{{Eur. Phys. J. Plus} {\bf 138}, 91 (2023)}.

\bibitem{tableviolation} V. A. Kosteleck\'y and N. Russell, ``Data tables for Lorentz and CPT violation,'' \doi{10.1103/RevModPhys.83.11}{{Rev. Mod. Phys.} {\bf 83}, 11 (2011)}.

\bibitem{electron1} Y. Ding and V. A. Kosteleck\'y, ``Lorentz-violating spinor electrodynamics and Penning traps,'' \doi{10.1103/PhysRevD.94.056008}{{Phys. Rev. D} {\bf 94}, 056008 (2016)}.

\bibitem{electron2} B. Altschul, ``Bounds on spin-dependent Lorentz violation from inverse Compton observations,'' \doi{10.1103/PhysRevD.75.041301}{{Phys. Rev. D} {\bf 75}, 041301 (2007)}.

\bibitem{electron3} B. Altschul, ``Synchrotron and inverse Compton constraints on Lorentz violations for electrons,'' \doi{10.1103/PhysRevD.74.083003}{{Phys. Rev. D} {\bf 74}, 083003 (2006)}.

\bibitem{protonformfactor} M. Gourdin, ``Electromagnetic form factors,'' \doi{10.1007/BF02750666} {{Nuovo Cim.} {\bf 36}, 129 (1965)}.

\bibitem{formfactorbook} S. Gasiorowicz, "Elementary particle physics," John Wiley \& Sons (1966).

\bibitem{protonformfactor2} F. J. Ernst, R. G. Sachs, and K. C. Wali, ``Electromagnetic Form Factors of the Nucleon,'' \doi{10.1103/PhysRev.119.1105}{{Phys. Rev.} {\bf 119}, 1105 (1960)}.

\bibitem{formfactors} M. Nowakowski, E. A. Paschos, and J. M. Rodriguez, ``All electromagnetic form factors,'' \doi{10.1088/0143-0807/26/4/001}{{Eur. J. Phys.,} {\bf 26}, 545 (2005)}.

\bibitem{edmexpression} M. Haghighat, I. Motie and Z Rezaei, ``Charged lepton electric dipole moment enhancement in the lorentz violated extension of the standard model,'' \doi{10.1142/S0217751X13501157}{{Int. J. Mod. Phys. A} {\bf 08}, 24 (2013)}.

\bibitem{edm01} D. Ghosh and R. Sato, ``Lepton electric dipole moment and strong CP violation,'' \doi{10.1016/j.physletb.2017.12.052}{{Phys. Lett. B} {\bf 777}, 335 (2018)}.

\bibitem{edm00} R. E. Rand, ``Determination of the Upper Limit to the Electric Dipole Moment of the Electron at High Momentum Transfer,'' \doi{10.1103/PhysRev.140.B1605}{{Phys. Rev.} {\bf 140}, B1605 (1965)}.

\bibitem{edm02} B. C. Regan, E. D. Commins, C. J. Schimdt and D. DeMille, ``New Limit on the Electron Electric Dipole Moment,'' \doi{10.1103/PhysRevLett.88.071805}{{Phys. Rev. Lett.} {\bf 88}, 071805 (2002)}

\bibitem{edm03} G. W. Bennet et al. ``Improved limit on the muon electric dipole moment,'' \doi{10.1103/PhysRevD.80.052008}{{Phys. Rev. D} {\bf 80}, 052008 (2009)}.

\bibitem{edm04} F. J. M. Farley, K. Jungmann, J. P. Miller, W. M. Morse, Y. F. Orlov, B. L. Roberts, Y. K. Semertzidis, A. Silenko, and E. J. Stephenson, ``New Method of Measuring Electric Dipole Moments in Storage Rings,'' \doi{10.1103/PhysRevLett.93.052001}{{Phys. Rev. Lett.} {\bf 93}, 052001 (2004)}.

\bibitem{edm05} W. Bernreuther, L. Chen and O. Nachtmann ``Electric dipole moment of the tau lepton revisited,'' \doi{10.1103/PhysRevD.103.096011}{{Phys. Rev. D} {\bf 103}, 096011 (2021)}.

\bibitem{edm06} K. S. Babu, B. Dutta and R. N. Mohapatra, ``Enhanced Electric Dipole Moment of the Muon in the Presence of Large Neutrino Mixing,'' \doi{10.1103/PhysRevLett.85.5064}{{Phys. Rev. Lett.} {\bf 85}, 5064 (2000)}. 

\bibitem{process} M. H. Sis, B. Mirza and A. K. B. Sefidi, ``$e^{-}e^{+}\to l^{-}l^{+}$ scattering in a strong magnetic field and LV background'', \doi{10.1016/j.aop.2022.169173}{{Ann. Phys.} {\bf 448}, 169173 (2023)}.

\bibitem{electronedm} ACME Collaboration, ``Improved limit on the electric dipole moment of the electron,'' \doi{10.1038/s41586-018-0599-8}{{Nature} {\bf 562}, 355 (2018)}.

\bibitem{edmweak} R. Budny, B. Kayser and J. Primack, ``Electric- and weak magnetic-dipole-moment effects in $e^{+}e^{-}\to l^{+}l^{-}$,'' \doi{10.1103/PhysRevD.15.1222}{{Phys. Rev. D} {\bf 15}, 1222 (1997)}.

\bibitem{danielmag} D. S. Cabral and A. F. Santos, ``$e^{+}e^{-}\to l^{+}l^{-}$ scattering at finite temperature in the presence of a classical background magnetic field,'' \doi{10.1140/epjp/s13360-024-04975-w}{{Eur. Phys. J. Plus} {\bf 139}, 190 (2024)}.

\bibitem{mag01} Y. Kazama, C. N. Yang and A. S. Goldhaber, ``Scattering of a Dirac particle with charge $Ze$ by a fixed magnetic monopole,'' \doi{10.1103/PhysRevD.15.2287}{{Phys. Rev. D} {\bf 15}, 2287 (1997)}.

\bibitem{mag02} L. Fassio-Canuto, ``Neutron Beta Decay in a Strong Magnetic Field,'' \doi{10.1103/PhysRev.187.2141}{{Phys. Rev.} {\bf 187}, 2141 (1969)}.

\bibitem{mag03} D. B. Melrose and A. J. Parle, ``Quantum Electrodynamics in Strong Magnetic Fields. I. Electron States,'' \doi{10.1071/PH830755}{{Aust. J. Phys.} {\bf 36}, 755 (1983)}.

\bibitem{paulicurrent} M. Nowakowski, ``The quantum mechanical current of the Pauli equation,'' \doi{10.1119/1.19149} {{Am. J. Phys.,} {\bf 67}, 916 (1999)}.

\bibitem{anomalous1} S. D. Drell and H. R. Pagels, ``Anomalous Magnetic Moment of the Electron, Muon and Nucleon,'' \doi{10.1103/PhysRev.140.B397}{{Phys. Rev.}  {\bf 140}, B397 (1965)}.

\bibitem{anomalous2} S. Narison, ``The anomalous magnetic moment of a charged heavy lepton,'' \doi{/10.1088/0305-4616/4/12/006}{{J. Phys. G: Nucl. Phys.,} {\bf 4}, 1849 (1978)}.

\bibitem{anomalous3} G. W. Bennett et al., ``Measurement of the Positive Muon Anomalous Magnetic Moment to 0.7 ppm,'' \doi{10.1103/PhysRevLett.89.101804}{{Phys. Rev. Lett.} {\bf 89}, 101804 (2002)}.

\bibitem{anomalous4} B. Abi et al., ``Measurement of the Positive Muon Anomalous Magnetic Moment to 0.46 ppm,'' \doi{10.1103/PhysRevLett.126.141801}{{Phys. Rev. Lett.} {\bf 126}, 141801 (2021)}.

\bibitem{anomalous5} A. Crivellin, F. Kirk and M. Schreck, ``Impact of Lorentz violation on anomalous magnetic moments of charged leptons,'' \doi{10.1007/JHEP11(2022)109} {{Journal of High Energy Physics} {\bf 109}, 2022 (2022)}.

\bibitem{anomalous6} R. Bluhm, V. A. Kostelecky and C. D. Lane, ``CPT and Lorentz Tests with muons,'' \doi{10.1103/PhysRevLett.84.1098} {{Phys. Rev. Lett.} {\bf 84}, 1098 (2000)}.

\bibitem{anomalous7} R. Bluhm, V. A. Kostelecky and N. Russell, ``Testing CPT with Anomalous Magnetic Moments,'' \doi{10.1103/PhysRevLett.79.1432}{{Phys. Rev. Lett.} {\bf 79}, 1432 (1997)}. 

\bibitem{zel} Ia B. Zel\'dovich, ``Electromagnetic interaction with parity violation,'' {{Sov. Phys. JETP} {\bf 6}, 1184 (1958)}.

\bibitem{anapoleweak} N. Dombey and A. D. Kennedy, ``A calculation of the electron anapole moment,'' \doi{10.1016/0370-2693(80)91013-8}{{Phys. Lett. B} {\bf 91}, 428 (1980)}.

\bibitem{violacao1kostelecky} D. Colladay and V. A. Kosteleck\'{y}, ``Cross sections and Lorentz violation,'' \doi{10.1016/S0370-2693(01)00649-9}{{Phys. Lett. B} {\bf 511}, 209 (2001).}

\bibitem{eliminatef} B. Altschul, ``Eliminating the CPT-odd f coefficient from the Lorentz-violating standard model extension,'' \doi{10.1088/0305-4470/39/44/010} {{J. Phys. A: Math. Gen.}  {\bf 39}, 13757 (2006)}.

\bibitem{magfield} K. Bhattacharya, ``Solution of the Dirac equation in presence of an uniform magnetic field,'' \arXivOld{0705.4275}.

\bibitem{magfieldthesis} K. Bhattacharya, ``Elementary particle interactions in a background magnetic field,'' \arXivOld{hep-ph/0407099}.

\bibitem{perturbation1} A. H. Nayfeh, ``Perturbation methods,'' John Wiley \& Sons (2008).

\bibitem{perturbation2} A. H. Nayfeh, ``Introduction to perturbation techniques,'' John Wiley \& Sons (2011).

\bibitem{perturbation3} B. K. Shivamoggi, ``Perturbation methods for differential equations,'' Boston: Birkh\"auser (2003).

\bibitem{field1} S. K. Lander and D. I. Jones, ``Magnetic fields in axisymmetric neutron stars,'' \doi{10.1111/j.1365-2966.2009.14667.x}{{Monthly Notices of the Royal Astronomical Society} {\bf 395}, 4 (2009)}.

\bibitem{field2} C. Y. Cardall, M. Prakash adn J. M. Lattimer, ``Effects of Strong Magnetic Fields on Neutron Star Structure,'' \doi{10.1086/321370}{{APJ} {\bf 554}, 322 (2001)}.

\bibitem{chiralbook} M. Srednicki, ``Quantum field theory,'' Cambridge University Press (2007).

\bibitem{magresult} A. Tiwari and B. K. Patra, ``Lowest-order electron-electron and electron-muon scattering in a strong magnetic field,'' \arXivOld{1808.04236}.


\end{thebibliography}
\end{document}